\documentclass{article}
\usepackage{amsmath}
\usepackage{array}
\usepackage{amsfonts}
\usepackage{amsmath}
\usepackage{rotate,epsfig}
\usepackage{amsfonts}
 \usepackage{lscape}
 \DeclareMathSizes{1}{5}{5}{5}
\newcommand{\iix}[1]{\mbox{$\displaystyle \int_{-\infty}^{\infty} #1 dx$}}

\newcommand{\prntz}[1]{\mbox{$\displaystyle \left( #1 \right)$}}
\newcommand{\xr}{\mbox{$\displaystyle X_{r:k}$}}
\newcommand{\lddp}[1]{\mbox{$\displaystyle \log\prntz{\frac{d F^{-1}(p)}{d p}}$}}

\newcommand{\xrr}{\mbox{$\displaystyle X_{r:k}^{(r)}$}}

\begin{document}

\title{Improved estimator of the entropy and goodness of fit tests in ranked set sampling}
\author{Morteza Amini\footnote{Corresponding author.  \newline \textit{E-mail
addresses:} a2odd1@yahoo.com (M. Mehdizadeh), mort.amini@gmail.com
(M. Amini), arghami@math.um.ac.ir(N. R. Arghami). }, M. Mehdizadeh
and N. R. Arghami\\
{\small {\it  Department of statistics, School of Mathematical
Sciences, Ferdowsi University of Mashhad,}}\vspace{-0.2cm}\\
{\small {\it P.O. Box 91775-1159,  Mashhad,\
Iran}}\vspace{-0.2cm} }

 \maketitle

\begin{abstract}
The entropy is one of the most applicable uncertainty measures in
many statistical and engineering problems. In statistical
literature, the entropy is used in calculation of the
Kullback-Leibler (KL) information which is a powerful mean for
performing goodness of fit tests. Ranked Set Sampling (RSS) seems
to provide improved estimators of many parameters of the
population in the huge studied problems in the literature. It is
developed for situations where the variable of interest is
difficult or expensive to measure, but where ranking in small
sub-samples is easy. In This paper, we introduced two estimators
for the entropy and compare them with each other and the
estimator of the entropy in Simple Random Sampling (SRS) in the
sense of bias and Root of Mean Square Errors (RMSE). It is
observed that the RSS scheme would improve this estimator. The
best estimator of the entropy is used along with the estimator of
the mean and two biased and unbiased estimators of variance based
on RSS scheme, to estimate the KL information and perform goodness
of fit tests for exponentiality and normality. The desired
critical values and powers are calculated. It is also observed
that RSS estimators would increase powers.
\end{abstract}


\section{Introduction}
\indent Suppose a continuous random variable $X$ has cumulative
distribution function (cdf) $F(x)$ and a probability density
function (pdf) $f(x)$. The differential entropy $H(f)$ of the
random variable $X$ is defined to be
\begin{equation}
H(f)=-\iix{f(x)\log f(x)}.
\end{equation}
The entropy is one of the most applicable uncertainty measures in
many statistical and engineering problems. In statistical
literature, the entropy is used in calculation of the
Kullback-Leibler (KL) information which is a powerful mean for
performing goodness of fit tests. The Kullback-Leibler (K-L)
information of $f(x)$ against $f_0(x)$ is defined in [7] to be
\begin{equation}
I(f;f_0)=\iix{f(x)\log \frac{f(x)}{f_0(x)}}.
\end{equation}
Since $I(f;f_0)$ has the property that $I(f;f_0)\geq0$, and the
equality holds if only if $f=f_0$, the estimate of the K-L
information has also been considered as a goodness of fit test
statistic by some authors including [2] and [5]. It has been
shown in the aforementioned papers that the test statistics based
on the K-L information perform very well for testing
exponentiality [5] as compared, in terms of power, with some
leading test statistics.

Ranked Set Sampling (RSS) has been developed by McIntyre (1952).
This method is applied for situations in which measuring a
variable is costly or difficult, but where ranking in small
subsets is easy. In this method, we first subdivide a sample of
size $n=k^2$ randomly into $k$ subsamples of size $k$, rank each
subsample visually or using any simple or cheap method and then
in the $r^{\rm th}$ subsample, measure and record only the unit of
rank $r$ which is denoted by $\xrr$ $(r=1,\ldots,k)$. Since the
subsamples are independent, $\xrr$'s are independent random
variables. Also the marginal distribution of $\xrr$ is the same
as that of $r^{\mbox{\rm th}}$ order statistic from a sample of
size $k$ of $X$, i.e. $\xr$. As it was proved by McIntyre, mean of
this sample is an unbiased estimator of the mean of $Y$ with an
efficiency slightly less than $\frac{1}{2}(k+1)$, relative to the
mean of a Simple Random Sample (SRS) of size $k$. Thus ``ranked
set sampling should be useful when the quantification of an
element is difficult but the elements of a set are easily drawn
and ranked by judgment." (Dell and Clutter 1972).

This method was also extended to estimating variance (Stokes
1980a), correlation coefficient (Stokes 1980b) and the situations
in which the sample is subdivided into subsamples of different
sizes.

In This paper, we introduced two estimators for the entropy and
compare them with each other and the estimator of the entropy in
Simple Random Sampling (SRS) in the sense of bias and Root of
Mean Square Errors (RMSE). It is observed that the RSS scheme
would improve this estimator. The best estimator of the entropy
is used along with the estimator of the mean and two biased and
unbiased estimators of variance based on RSS scheme, to estimate
the KL information and perform goodness of fit tests for
exponentiality and normality. The desired critical values and
powers are calculated. It is also observed that RSS estimators
would increase powers.

\section{Entropy estimation}

The nonparametric estimation of the entropy
\begin{equation}\label{H}
H=\int_{0}^{1}\log \left(\frac{dF^{-1}(p)}{dp}\right)dp.
\end{equation}
 An estimate of (\ref{H}) can be constructed by replacing the distribution
function $F$ by the empirical distribution $F_{n}.$ The
derivative of $F^{-1}(i/n)$ is estimated by
$(x_{i+w:n}-x_{i-w:n})n/(2w)$. The estimate of $H$ is then
\begin{equation}
H(m,n)=\frac{1}{n}\sum_{i=1}^{n}\log
\left(\frac{n}{2m}(x_{i+m:n}-x_{i-m:n})\right),
\end{equation}

\begin{table}
  \centering
  \caption{Simulated Minimum RMSE (MRMSE) and Minimum Absolute Bias (MAB) of $H_{mn}^1$ and $H_{mn}^2$ and optimal
  $m$ for $k=10$ and three distributions with different values of $r$.
  }\label{twoest}
\begin{tabular}{c c c c c c} \hline
&  &  & $r$ &  &  \\
& & 2& & 3 &  \\
\hline
&   &$H_{mn}^1$&$H_{mn}^2$&$H_{mn}^1$&$H_{mn}^2$\\
 \hline
U(0,1) & MRMSE  (optimal $m$*)   & 0.062(8) & 0.081(5) & 0.045(11-13) & 0.073(5) \\
     & MAB(optimal $m$*)         & 0.030(10) & 0.047(5) & 0.021(15)   & 0.048(5) \\
\hline
e(1) & MRMSE  (optimal $m$*)     & 0.157(5) & 0.168(4)  & 0.125(6) & 0.140(4) \\
     & MAB(optimal $m$*)         & 0.001(6) & 0.0137(5) & 0.004(7) & 0.014(5) \\
\hline
N(0,1) & MRMSE  (optimal $m$*)   & 0.184(5,10) & 0.246(5) & 0.138(7,8) & 0.233(5) \\
     & MAB(optimal $m$*)         & 0.113(10) & 0.205(5) & 0.062(12) &  0.206(5)\\
\hline
\end{tabular}\\
*{\small $m=1(1)k/2$ for $H_{mn}^2$ and $m=1(1)rk/2$ for
$H_{mn}^1$}
\end{table}

where the window size $m$ is a positive integer, which is less
than $n/2$, and $x_{i:n}=x_{1:n}$ for $i<1$, and
$x_{i:n}=x_{n:n}$ for $i>n$.

Ebrahimi et al. (1994) proposed a modified sample entropy as
\begin{equation}
H_{c}(n,m)=n^{-1}\sum_{i=1}^{n}\log\frac{n}{c_{i}m}(X_{(i+m)}-X_{(i-m)})
\end{equation}
where
\begin{displaymath}
c_{i}=\left\{ \begin{array}{ll}
       1+\frac{i-1}{m} & \textrm{if $1\leq i \leq m$}\\
       2 & \textrm{if $m+1\leq i \leq n-m$}\\
       1+\frac{n-i}{m} & \textrm{if $n-m+1\leq i \leq n$}
       \end{array} \right.
\end{displaymath}

To estimate the entropy in RSS scheme, we may note that the
estimator of $F^{-1}(i/n)$ must be positive for $\log$ function
to be well-defined. So we have to order the ranked set sample.
There are two ways to order this sample. First way is to order
each replication, derive the estimator and then take the average
as the main estimator. The second way is to order the whole
sample of size $rk$. This two methods yield two estimators as
follows
\begin{equation}
H_{mn}^1=\frac{1}{n}\sum_{i=1}^{n}\log\frac{n}{c_{i}m}(X_{[i+m]}-X_{[i-m]})
\end{equation}
and
\begin{equation}
H_{mn}^2=\frac{1}{n}\sum_{j=1}^{r}\sum_{i=1}^{k}\log\frac{k}{d_{i}m}(X_{[i+m]j}-X_{[i-m]j}),
\end{equation}

\begin{landscape}
\begin{table}
  \centering
  \caption{\small Monte Carlo biases and RMSE for $H_{mn}^1$ in three distributions for
  $n=10,20$}\label{rmsebias1}
\begin{tabular}{c c | c c c c | c c c c | c c c c } \hline
& & & & & & & & & & & & & \\
& & & { U(0,1)}& & & & { e(1)} & & & & { N(0,1)} & & \\
 \cline{3-5}\cline{7-9}\cline{11-13}\\
   &   & { SRS}  &      &  { RSS}  &  & { SRS}  &      &  { RSS } & & { SRS } &      & { RSS}   &   \\
   \cline{3-14}\\
 { $n$ }& { $m$ }& { Bias }& { RMSE} &  { Bias }& { RMSE} &  { Bias }& { RMSE}&  { Bias }& { RMSE} & { Bias }& { RMSE} &  { Bias }& { RMSE} \\
\hline
10 & 1 & -0.381 & 0.451 & -0.259 & 0.326 & -0.392 & 0.561 & -0.298 & 0.398 &-0.452 & 0.458 & -0.342 & 0.428 \\
   & 2 & -0.222 & 0.293 & -0.108 & 0.168 & -0.222 & 0.436 & -0.142 & 0.266 &-0.342 & 0.441 & -0.227 & 0.311 \\
   & 3 & -0.159 & 0.228 & -0.070 & 0.124 & -0.174 & 0.405 & -0.078 & 0.241 &-0.301 & 0.408 & -0.207 & 0.289 \\
   & 4 & -0.140 & 0.224 & -0.056 & 0.110 & -0.114 & 0.382 & -0.031 & 0.236 &-0.305 & 0.394 & -0.209 & 0.285 \\
   & 5 & -0.131 & 0.212 & -0.050 & 0.107 & -0.064 & 0.371 &  0.012 & 0.249 &-0.289 & 0.389 & -0.204 & 0.279 \\
& & & & & & & & & & & & & \\
20* & 1 & -0.328 & 0.358 & -0.274 & 0.302 & -0.335 & 0.424 & -0.290 & 0.340 &-0.373  & 0.427 & -0.313 & 0.358\\
   & 2 & -0.176 & 0.203 & -0.121 & 0.147 & -0.179 & 0.299 & -0.139 & 0.206 & -0.221 & 0.288 & -0.178 & 0.232\\
   & 3 & -0.125 & 0.155 & -0.076 & 0.103 & -0.151 & 0.280 & -0.083 & 0.169 & -0.179 & 0.252 & -0.141 & 0.200\\
   & 4 & -0.104 & 0.134 & -0.056 & 0.084 & -0.098 & 0.264 & -0.052 & 0.161 & -0.176 & 0.255 & -0.124 & 0.189\\
   & 5 & -0.088 & 0.119 & -0.046 & 0.074 & -0.062 & 0.253 & -0.024 & 0.157 &-0.167  & 0.245 & -0.117 & 0.184\\
   & 6 & -0.079 & 0.117 & -0.040 & 0.067 & -0.047 & 0.244 & 0.001 & 0.165 & -0.156  & 0.232 & -0.116 & 0.185\\
   & 7 & -0.076 & 0.111 & -0.035 & 0.063 & -0.020 & 0.263 & 0.025 & 0.173 &-0.150  & 0.231  & -0.116 & 0.185\\
   & 8 & -0.068 & 0.109 & -0.034 & 0.062 &  0.010 & 0.264 & 0.051 & 0.188 &-0.157  & 0.239  & -0.114 & 0.185\\
   & 9 & -0.064 & 0.108 & -0.032 & 0.063 &  0.032 & 0.260 & 0.078 & 0.203 &-0.158  & 0.241  & -0.116 & 0.188\\
   & 10& -0.061 & 0.106 & -0.030 & 0.063 &  0.044 & 0.268 & 0.102 & 0.225 & -0.152 & 0.234  & -0.113 & 0.184\\
\hline
\end{tabular}\\
\hspace{-7cm}{*{\small $n=10r$ cases are observed by RSS scheme
with 10 samples and $r$ replication.}}
\end{table}
\end{landscape}
\newpage

\begin{landscape}
\begin{table}
  \centering
  \caption{\small Monte Carlo biases and RMSE for $H_{mn}^1$ in three distributions for
  $n=30$}\label{rmsebias2}
\begin{tabular}{c c | c c c c | c c c c | c c c c } \hline
& & & & & & & & & & & & & \\
& & & { U(0,1)}& & & & { e(1)} & & & & { N(0,1)} & & \\
 \cline{3-5}\cline{7-9}\cline{11-13}\\
   &   & { SRS}  &      &  { RSS}  &  & { SRS}  &      &  { RSS } & & { SRS } &      & { RSS}   &   \\
   \cline{3-14}\\
 { $n$ }& { $m$ }& { Bias }& { RMSE} &  { Bias }& { RMSE} &  { Bias }& { RMSE}&  { Bias }& { RMSE} & { Bias }& { RMSE} &  { Bias }& { RMSE} \\
\hline
30* & 1 & -0.312 & 0.337 & -0.273 & 0.290 & -0.293 & 0.363 & -0.286 & 0.318 &-0.328 & 0.370 & -0.300 & 0.331 \\
   & 2 & -0.158 & 0.174 & -0.125 & 0.141 & -0.156 & 0.248 & -0.136 & 0.181 & -0.198 & 0.245 & -0.160 & 0.199\\
   & 3 & -0.110 & 0.129 & -0.080 & 0.096 & -0.115 & 0.219 & -0.084 & 0.145 & -0.158 & 0.209 & -0.118 & 0.164\\
   & 4 & -0.090 & 0.110 & -0.058 & 0.075 & -0.078 & 0.206 & -0.052 & 0.129 & -0.135 & 0.196 & -0.099 & 0.150 \\
   & 5 & -0.071 & 0.092 & -0.046 & 0.065 & -0.059 & 0.202 & -0.034 & 0.126 & -0.113 & 0.184 & -0.088 & 0.144\\
   & 6 & -0.069 & 0.090 & -0.039 & 0.058 & -0.045 & 0.198 & -0.013 & 0.125 & -0.106 & 0.181 & -0.082 & 0.142\\
   & 7 & -0.061 & 0.083 & -0.034 & 0.054 & -0.023 & 0.206 & 0.004 & 0.129 & -0.098 & 0.174 & -0.073 & 0.138\\
   & 8 & -0.056 & 0.079 & -0.030 & 0.050 & -0.007 & 0.194 & 0.021 & 0.135 & -0.106 & 0.175 & -0.069 & 0.138\\
   & 9 & -0.052 & 0.076 & -0.028 & 0.048 &  0.015 & 0.192 & 0.039 & 0.145 & -0.086 & 0.174 & -0.067 & 0.140\\
   & 10& -0.050 & 0.075 & -0.027 & 0.046 &  0.027 & 0.195 & 0.057 & 0.155 & -0.091 & 0.175 & -0.067 & 0.141\\
   & 11& -0.048 & 0.075 & -0.025 & 0.045 &  0.050 & 0.211 & 0.073 & 0.167 & -0.091 & 0.171 & -0.065 & 0.141 \\
   & 12& -0.041 & 0.071 & -0.024 & 0.045 &  0.075 & 0.225 & 0.098 & 0.185 & -0.090 & 0.171 & -0.062 & 0.141\\
   & 13& -0.042 & 0.074 & -0.023 & 0.045 &  0.089 & 0.231 & 0.117 & 0.201 & -0.089 & 0.175 & -0.065 & 0.144 \\
   & 14& -0.043 & 0.073 & -0.022 & 0.046 &  0.100 & 0.248 & 0.132 & 0.214 & -0.090 & 0.174 & -0.066 & 0.143 \\
   & 15& -0.037 & 0.069 & -0.021 & 0.046 &  0.124 & 0.255 & 0.150 & 0.231 & -0.094 & 0.177 & -0.064 & 0.143\\
\hline
\end{tabular}\\
\hspace{-7cm}{*{\small $n=10r$ cases are observed by RSS scheme
with 10 samples and $r$ replication.}}
\end{table}
\end{landscape}

where
\begin{displaymath}
d_{i}=\left\{ \begin{array}{ll}
       1+\frac{i-1}{m} & \textrm{if $1\leq i \leq m$}\\
       2 & \textrm{if $m+1\leq i \leq k-m$}\\
       1+\frac{k-i}{m} & \textrm{if $k-m+1\leq i \leq k$}
       \end{array} \right..
\end{displaymath}

Table \ref{twoest} shows the values of simulated Minimum RMSE
(MRMSE) and Minimum Absolute Bias (MAB) of $H_{mn}^1$ and
$H_{mn}^2$ and optimal $m$ for $k=10$ and three famous
distributions with different values of $r$. From this values one
can conclude that $H_{mn}^1$ is better estimator in the sense of
RMSE and bias. Tables \ref{rmsebias1} and \ref{rmsebias2} show
the values of Monte Carlo biases and RMSE for $H_{mn}^1$ in three
distributions for $n=10,20$ and 30. This values present a
distinct improvement of the estimator in RSS scheme relative to
SRS scheme.

\section{Goodness of fit tests}

Park, S. and D. (2003) derived the nonparametric distribution
function of $H_{c}(n,m)$ as
\begin{displaymath}
g_{c}(x)=\left\{ \begin{array}{ll}
       0 & \textrm{if $x<\eta_{1}\quad$ or $\quad x>\eta_{n+1}$}\\
       n^{-1}\frac{1}{\eta_{i+1}-\eta_{i}} & \textrm{if $\eta_{i}<x\leq\eta_{i+1},i=1,...,n$}
       \end{array} \right.,
\end{displaymath}
where
\begin{displaymath}
\eta_{i}=\left\{ \begin{array}{ll}
       \xi_{m+1}-\sum_{k=i}^{m}\frac{1}{m+k-1}(x_{(m+k)}-x_{(1)}) & \textrm{if $1\leq i \leq m$}\\
       \frac{1}{2m}(x_{(i-m)}+...+x_{(i+m-1)}) & \textrm{if $m+1\leq i \leq n-m+1$}\\
       \xi_{n-m+1}+\sum_{k=n-m+2}^{i}\frac{1}{n+m-k+1}(x_{(n)}-x_{(k-m-1)}) & \textrm{if $n-m+2\leq i \leq n+1$}
       \end{array} \right.
\end{displaymath}
They used it to correct the moments of the distribution which are
used in goodness of fit tests.

In the exponentiality test, the aforementioned nonparametric
distribution is used to estimate the mean and
${\hat{\lambda}}_{c}$.

\begin{equation}
I(g:f)=\iix{g(x)\ln\frac{g(x)}{f(x)}}
\end{equation}

\begin{equation}
T_{c}=1+\log{\hat{\lambda}}_{c}-H_{c}(n,m)
\end{equation}

The following alternatives of the exponentiality null hypothesis
have been considered to estimate the powers.
\begin{description}
\item 1. Gamma distribution with pdf
\begin{equation}
f(x;\alpha)=\frac{x^{\alpha-1}\exp(-x)}{\gamma(\alpha)}\quad
\alpha>0,x>0
\end{equation}
\item 2. Weibull distribution with pdf
\begin{equation}
f(x;\beta)=\beta x^{\beta-1}\exp(-x^{\beta})\quad \beta>0,x>0
\end{equation}
\item 3. Log-normal distribution with pdf
\begin{equation}
f(x;\alpha)=\frac{1}{\sigma
\sqrt{2\pi}x}\exp(-\frac{1}{2{\sigma}^2}{(\log x)}^2)\quad
\sigma>0,x>0
\end{equation}
\item 4. Uniform distribution with pdf
\begin{equation}
f(x)=1 \quad 0<x<1
\end{equation}
\end{description}

As mentioned by Arizono and Ohta (1989), an estimate for
$I(f,f_0)$, when $f_0$ is the normal pdf with known parameters
$\mu$ and $\sigma$ is obtained as
\begin{equation}\label{imn}
I_{mn}=\log(\sqrt{2\pi\sigma^2})+\frac{1}{2n}\sum_{i=1}^{n}\prntz{\frac{x_i-\mu}{\sigma}}^2-H(n,m).
\end{equation}

\begin{table}
  \centering
  \caption{\small Critical values for different values of $n$, $m$ and $\alpha$}\label{crits}
\begin{tabular}{c c c c c c  c c c c c c c  }
 { Exponentiality}  &  &  & &   & &  & &  &  & & { Normality}  \\
\end{tabular}
\begin{tabular}{c c c c c c | c c c c } \hline
   \hline
   &  &  &$\alpha$ & &    &  &$\alpha$ &   \\
   \cline{3-6}\cline{7-10}
 { $n$ }& $m$ & { 0.1 }& { 0.05 } & { 0.025 }& { 0.01 } & { 0.1 }& { 0.05 } & { 0.025 }& { 0.01 } \\
\hline
10 & 1 & 0.5357 & 0.6318 & 0.7297 &  0.8617 &0.5898 & 0.7027  & 0.8034 &0.9215 \\
   & 2 & 0.2898 & 0.3546 & 0.4213 &  0.5099 &0.3765 & 0.4404  & 0.5113 &0.6005  \\
   & 3 & 0.2095 & 0.2645 & 0.3243 &  0.3944 &0.3214 & 0.3712  & 0.4182 &0.4667  \\
   & 4 & 0.1619 & 0.2154 & 0.2596 &  0.3293 &0.3001 & 0.3221  & 0.3593 &0.3987  \\
   & 5 & 0.1416 & 0.1916 & 0.2487 &  0.3122 &0.2903 & 0.3091  & 0.3311 &0.3544  \\
& & & & & & & & \\
20* & 1 & 0.4455 & 0.5091 & 0.5695 &  0.6373& 0.4775 & 0.5405 & 0.6025 &0.6587  \\
   & 2 & 0.2391 & 0.2822 & 0.3305 &  0.3813 & 0.2824 & 0.3264 & 0.3621 &0.4092  \\
   & 3 & 0.1707 & 0.2089 & 0.2450 &  0.2939 & 0.2296 & 0.2614 & 0.2940 &0.3460  \\
   & 4 & 0.1389 & 0.1738 & 0.2064 &  0.2498 & 0.2073 & 0.2339 & 0.2671 &0.3112  \\
   & 5 & 0.1117 & 0.1445 & 0.1772 &  0.2173 & 0.2000 & 0.2287 & 0.2549 &0.2875  \\
   & 6 & 0.0964 & 0.1269 & 0.1569 &  0.1918 & 0.1977 & 0.2255 & 0.2501 &0.2802  \\
   & 7 & 0.0779 & 0.1114 & 0.1441 &  0.1741 & 0.1968 & 0.2223 & 0.2463 &0.2695  \\
   & 8 & 0.0643 & 0.0983 & 0.1250 &  0.1754 & 0.2013 & 0.2225 & 0.2407 &0.2612  \\
   & 9 & 0.0495 & 0.0915 & 0.1188 &  0.1604 & 0.2023 & 0.2213 & 0.2375 &0.2569  \\
   & 10& 0.0368 & 0.0797 & 0.1167 &  0.1543 & 0.2008 & 0.2175 & 0.2391 &0.2512  \\
& & & & & & & &\\
30* & 1 & 0.4100 & 0.4567 & 0.4961 &  0.5656& 0.4273 & 0.4729 & 0.5155 &0.5768  \\
   & 2 & 0.2171 & 0.2498 & 0.2796 &  0.3156 & 0.2443 & 0.2776 & 0.3028 &0.3451  \\
   & 3 & 0.1516 & 0.1819 & 0.2122 &  0.2402 & 0.1891 & 0.2145 & 0.2408 &0.2777  \\
   & 4 & 0.1202 & 0.1481 & 0.1693 &  0.2065 & 0.1649 & 0.1855 & 0.2099 &0.2407  \\
   & 5 & 0.0979 & 0.1210 & 0.1503 &  0.1860 & 0.1498 & 0.1739 & 0.1912 &0.2300  \\
   & 6 & 0.0825 & 0.1102 & 0.1361 &  0.1559 & 0.1454 & 0.1658 & 0.1879 &0.2208  \\
   & 7 & 0.0722 & 0.0950 & 0.1212 &  0.1501 & 0.1433 & 0.1660 & 0.1891 &0.2134  \\
   & 8 & 0.0574 & 0.0849 & 0.1061 &  0.1449 & 0.1425 & 0.1663 & 0.1848 &0.2082  \\
   & 9 & 0.0574 & 0.0849 & 0.0960 &  0.1270 & 0.1453 & 0.1631 & 0.1833 &0.2046  \\
   & 10& 0.0379 & 0.0635 & 0.0878 &  0.1162 & 0.1428 & 0.1654 & 0.1838 &0.2039  \\
   & 11& 0.0280 & 0.0545 & 0.0760 &  0.1097 & 0.1468 & 0.1661 & 0.1838 &0.2056  \\
   & 12& 0.0151 & 0.0447 & 0.0706 &  0.1030 & 0.1489 & 0.1697 & 0.1875 &0.2049  \\
   & 13& 0.0043 & 0.0354 & 0.0640 &  0.0927 & 0.1502 & 0.1719 & 0.1891 &0.2072  \\
   & 14& -0.0046 & 0.0274 & 0.0612 &  0.0882& 0.1527 & 0.1720 & 0.1857 &0.2073  \\
   & 15& -0.0190 & 0.0182 & 0.0505 &  0.0813& 0.1492 & 0.1716 & 0.1871 &0.2091  \\
\hline
\end{tabular}
*{\small $n=10r$ cases are observed by RSS scheme with 10 samples
and $r$ replication.}

\end{table}

When both $\mu$ and $\sigma$ are unknown, we place their
estimates, that is, $\hat{\mu}=\bar{X}$ and
$\hat{\sigma}=\frac{1}{n}\sum_{i=1}^{n}(X_i-\bar{X})^2$ in
(\ref{imn}) and derive the test statistic as

\begin{equation}
T=\log(\sqrt{2\pi {\hat{\sigma}}^{2}})+0.5-H(n,m)
\end{equation}

Park, S. and D. replaced the estimates $H(n,m)$ and
$\hat{\sigma}$ with their corrected estimators $H_{c}(n,m)$ and
${\hat{\sigma}}_{c}$ and derived the test statistic
\begin{equation}
T_{c}=\log(\sqrt{2\pi {\hat{\sigma}}_{c}^{2}})+0.5-H_{c}(n,m)
\end{equation}

In the normality test the following alternatives are considered
to estimate the powers
\begin{description}
\item 1. Uniform distribution with pdf
\begin{equation}
f(x)=1 \quad 0<x<1
\end{equation}
\item 2. Chi-square distribution with pdf
\begin{equation}
f(x;\alpha)=\frac{1}{\Gamma(\alpha/2)}(\frac{1}{2})^{\alpha/2}x^{(\alpha/2)-1}\exp(-\frac{1}{2}x)\quad
\alpha>0,x>0
\end{equation}
\item 3. t-student distribution with pdf
\begin{equation}
f(x;\nu)=\frac{\Gamma((\nu+1)/2)}{\Gamma(\nu/2)}\frac{1}{\sqrt(\nu\pi)}\frac{1}{(1+x^{2})^{(\nu+1)/2}}\quad
\nu>2,-\infty<x<\infty
\end{equation}
\item 4. Exponential distribution with pdf
\begin{equation}
f(x;\lambda)=\lambda\exp(-\lambda x),\quad\lambda>0,\;x>0.
\end{equation}
\end{description}

Stokes (1980) proposed the sample variance as an estimator of the
population variance as follows
\begin{equation}\label{stks}
{\hat{\sigma}}^{2}=\frac{1}{rk-1}\sum_{i=1}^{r}\sum_{j=1}^{k}(X_{[j]i}-\hat{\mu})^{2}
\end{equation}
This estimator is asymptotically unbiased and asymptotically more
efficient than the sample variance in SRS. MacEachern et al.
(2002) proposed an unbiased estimator of the variance as follows
\begin{equation}
{\tilde{\sigma}}^{2}=\frac{1}{rk}{(k-1)\textrm{MST}}+(rk-k+1)\textrm{MSE},
\end{equation}
where
\begin{equation}
\textrm{MST}=\frac{1}{k-1}\sum_{i}\sum_{j}{(X_{j[i]}-
\hat{\mu})}^2-\frac{1}{k-1}\sum_{j}\sum_{i}{(X_{j[i]}-{\bar{X}}_{[j].})}^{2},
\end{equation}

\begin{equation}
\textrm{MSE}=\frac{1}{k(r-1)}\sum_{j}\sum_{i}{(X_{[j]i}-{\bar{X}}_{[j].})}^{2},
\end{equation}

\begin{table}
  \centering
  \caption{\small Power comparison of 0.05 tests against some alternatives in SRS and RSS schemes}\label{comp}
\begin{tabular}{ c c c c c c c c c c c c }
&    &  &   &   & Exponentiality &    &  &   &   & \\
\end{tabular}
\begin{tabular}{ c c c c c c c c } \hline
  &  &    &  & $n$  &   &   &  \\
 \cline{3-7}
  &  & 20* &  &   &   &  50*  &  \\
  &  & {\tiny($m=4$)} &  &   &   & {\tiny($m=6$)} &  \\
\cline{2-4}\cline{6-8}
  { Alternatives }& { SRS}&& { RSS} &&  { SRS }&& { RSS}  \\
\hline
    Gamma (1.5)   & 0.2176 &&  0.2740  && 0.3480 &&  0.4193  \\
    Lognormal (1) & 0.2685 &&  0.1908  && 0.6613 &&  0.4156  \\
    Weibull (1.5) & 0.4639 &&  0.6199  && 0.7752 &&  0.9056  \\
    Gamma (2)     & 0.4862 &&  0.6218  && 0.8281 &&  0.9050  \\
    Gamma (3)     & 0.8816 &&  0.9693  && 0.9993 &&  0.9999  \\
    Uniform       & 0.8021 &&  0.9979  && 0.9989 &&  1.0000  \\
    Weibull (2)   & 0.9138 &&  0.9896  && 0.9995 &&  1.0000   \\
  Lognormal (0.5) & 0.9967 &&  0.9994  && 1.0000 &&  1.0000  \\
  \hline
  Average power   & 0.6288 &&  0.7078  && 0.8263 &&  0.8307  \\
\hline
\end{tabular}
\begin{tabular}{ c c c c c c c c c c c c }
&    &  &   &   & Normality &    &  &   &   & \\
\end{tabular}
\begin{tabular}{ c c c c c c c c } \hline
  &  &    &  & $n$  &   &   &  \\
 \cline{3-7}
  &  & 20* &  &   &   &  50*  &  \\
  &  & {\tiny($m=3$)} &  &   &   & {\tiny($m=4$)} &  \\
\cline{2-4}\cline{6-8}
  { Alternatives }& { SRS}& { RSS} &  &&  { SRS }&{ RSS} &   \\
  \cline{2-4}\cline{6-8}
        &       & {\small $KL_{mn}^2$}  & {\small $KL_{mn}^1$}  &&  &  {\small $KL_{mn}^2$}  & {\small $KL_{mn}^1$} \\
\hline
   t(5)                               & 0.1069 & 0.0847 & 0.0865  && 0.2395 &0.1515& 0.1497   \\
   t(3)                               & 0.1989 & 0.1761 & 0.1748  && 0.5132 &0.4009& 0.3868   \\
   Uniform                            & 0.3851 & 0.4801 & 0.4897  && 0.8850 &0.9843& 0.9800   \\
   $\chi^2_4$                         & 0.5058 & 0.5704 & 0.5739  && 0.9326 &0.9709& 0.9710   \\
   $\chi^2_2$ {\small (Exponential)}  & 0.8656 & 0.9574 & 0.9650  && 0.9997 &1.0000& 1.0000   \\
   $\chi^2_1$                         & 0.9934 & 0.9999 & 0.9999  && 1.0000 &1.0000& 1.0000   \\
   \hline
  Average power                       & 0.5093 & 0.5448 & 0.5483  && 0.5713 &0.7513&  0.7479  \\
\hline
\end{tabular}\\
{*{\small $n=10r$ cases are observed by RSS scheme with 10 samples
and $r$ replication.}}
\end{table}

\begin{equation}
{\bar{X}}_{[j].}=\sum_{i}X_{[j]i}/r.
\end{equation}
and
\begin{equation}\label{mh}
\hat{\mu}=\sum_{i}\sum_{j}X_{[j]i}/rk.
\end{equation}
If we use our entropy estimator for estimation of
Kullback-Leibler distance between an unknown pdf and the pdf of
the normal distribution, we derive
\begin{equation}\label{kmn}
K_{mn}=\log(\sqrt{2\pi\sigma^2})+\frac{1}{2n}\sum_{i=1}^{n}\prntz{\frac{x_i-\mu}{\sigma}}^2-H^2_{mn}.
\end{equation}
In goodness of fit test of normality when $\mu$ and $\sigma$ are
unknown we can place their estimators in the RSS scheme, i.e.
$\hat{\mu}$ in (\ref{mh}) and the Stokes estimator, (\ref{stks})
to derive the test statistic as
\begin{equation}\label{imn}
KL^1_{mn}=\log(\sqrt{2\pi\hat{\sigma}^2})+0.5-H^2_{mn}.
\end{equation}
If we place the MacEachern et al. estimator of variance in
(\ref{kmn}), we derive another test statistic as
\begin{equation}\label{imn}
KL^2_{mn}=\log(\sqrt{2\pi\tilde{\sigma}^2})+\frac{1}{2n}\sum_{i=1}^{n}\prntz{\frac{x_i-\hat{\mu}}{\tilde{\sigma}}}^2-H^2_{mn}.
\end{equation}

Table \ref{crits} contains critical values of exponentiality and
normality tests for different values of $n$, $m$ and $\alpha$.

Table \ref{comp} propose a comparison of powers in RSS and SRS
schemes, for exponentiality and normality tests. The SRS values
of powers are given from Park. S. and D. with the modified sample
entropy of Ebrahimi et al. and their modified estimators of
moments. We used the similar window size $m$ for the comparison
although our maximum powers may be obtained for different values
of $m$. For normality test two test statistics $KL_{mn}^1$ and
$KL_{mn}^2$ are compared in the sense of power. For $n=20$, using
the statistic $KL_{mn}^2$ cause less powers than $KL_{mn}^1$.
Although the average of powers of $KL_{mn}^2$ gets larger than
the average power of $KL_{mn}^1$ when $n$ increases to 50, but
the difference between this powers is ignorable. Since obtaining
the statistic $KL_{mn}^2$ is more complicated than $KL_{mn}^1$,
we prefer to use $KL_{mn}^1$ for the remaining of the study.

Table \ref{maxpow} shows the maximum powers and the maximal
window size, $m$ for $\alpha=0.05$ of exponentiality and
normality tests. Ebrahimi et al. (1992) used such maximality to
obtain some optimal window size $m$ for each $n$. Table
\ref{maxpow} shows that here this values of optimal $m$ differs
distinctly for different alternatives. In fact choosing an
optimal $m$ depends very closely to the alternative which is
unknown. So in this paper we use the average of powers for
considered alternatives as a measure to decide about the optimal
$m$. The values of average powers are tabulated in Table
\ref{avepow}. The authors believe that this values are more
useful for the experimenter who wants to perform a test, since he
is not aware about the alternative. Table \ref{optm} shows the
optimal $m$ and the maximum average powers for different values
of $n$ of exponentiality and normality tests.

\begin{table}
  \centering
  \caption{\small Maximum powers (maximal $m$) of 0.05 tests against some alternatives
   of the null hypothesis distributions}\label{maxpow}
\begin{tabular}{ c c c c c c c c c c c c }
&    &  &   &   & Exponentiality &    &  &   &   & \\
\end{tabular}
\begin{tabular}{ c c c c c c  } \hline
    &    &  & $n$  &   &     \\
 \cline{2-6}
                    & 10        & 20*       & 30*        &  40*       &  50*   \\
  { $Alternatives$ }& &&  &&   \\
\hline
    Gamma (1.5)     & 0.5760(5) & 0.3761(8) & 0.4126(12) & 0.4371(14) & 0.4569(7)\\
    Lognormal (1)   & 0.1333(2) & 0.2140(3) & 0.3133(3) & 0.4143(4) & 0.5174(3)\\
    Weibull (1.5)   & 0.5999(5) & 0.7600(8) & 0.8365(15) & 0.8725(8) & 0.9099(7)\\
    Gamma (2)       & 0.5638(4) & 0.7216(8) & 0.7939(7) & 0.8570(8) & 0.9153(7)\\
    Gamma (3)       & 0.9023(5) & 0.9745(5) & 0.9956(5) & 0.9997(5) & 1.0000(3-5)\\
    Uniform         & 0.9201(5) & 1.0000(8,10) & 1.0000(3-15) & 1.0000(2-20) & 1.0000(2-25)\\
    Weibull (2)     & 0.9659(5) & 0.9963(8) & 0.9998(8,12) & 1.0000(4-12) & 1.0000(3-12)\\
    Lognormal (0.5) & 0.9815(4) & 0.9995(3) & 1.0000(2-6) & 1.0000(2-9) & 1.0000(2-12)\\
\hline
\end{tabular}
\begin{tabular}{ c c c c c c c c c c c c }
&    &  &   &   & Normality &    &  &   &   & \\
\end{tabular}
\begin{tabular}{ c c c c c c  } \hline
    &    &  & $n$  &   &     \\
 \cline{2-6}
                          & 10        & 20*         & 30*         &  40*           &  50*   \\
  { $Alternatives$ }& &&  &&   \\
\hline
    t(5)                  & 0.0813(4) & 0.0865(3)   & 0.1133(3)    & 0.1427(2)     & 0.1615(2) \\
    t(3)                  & 0.1335(4) & 0.1846(2)   & 0.2820(2)    & 0.3514(3)     & 0.4260(3) \\
    Uniform               & 0.1523(2) & 0.5805(10)  & 0.9036(11)   & 0.9875(16)    & 0.9992(16,20)\\
    $\chi^2_4$            & 0.3462(4) & 0.6164(4)   & 0.8305(6)    &  0.9334(6)    & 0.9781(5)\\
$\chi^2_2$ (Exponential)  & 0.6926(4) & 0.9670(4)   & 0.9992(4)    &  1.0000(3-8)  & 1.0000(1-14)\\
    $\chi^2_1$            & 0.9492(3) & 0.9999(3-5) & 1.0000(1-12) &  1.0000(1-17) & 1.0000(1-22)\\
\hline
\end{tabular}\\
{*{\small $n=10r$ cases are observed by RSS scheme with 10 samples
and $r$ replication.}}
\end{table}

\begin{table}
  \centering
  \caption{\small Average powers $\alpha=0.05$ for different alternatives and different values of $n$ and
  $m$}\label{avepow}
  \begin{tabular}{ c c c c c c c c c c c c c c c}
&    &  &   &  & & Exponentiality &    &  &   & &  & \\
\end{tabular}
\begin{tabular}{c c c c c c c c c c c c c c c }
\hline
 { $n$ }& $m$ & AP &  { $n$ }& $m$ & AP  &{ $n$ }& $m$ & AP &{ $n$ }& $m$ & AP&{ $n$ }& $m$ & AP\\
\hline
 10 & 1& 0.2905 & 30*& 1 & 0.5470& 40* & 1 & 0.6156&   40*& 16& 0.7692&  50* &11 & 0.7997\\
   & 2 & 0.5138 &    & 2 & 0.6851&     & 2 & 0.7325&      &17 & 0.7643&      & 12& 0.7950\\
   & 3 & 0.6281 &    & 3 & 0.7364&     & 3 & 0.7786&      & 18& 0.7634&      & 13& 0.7872\\
   & 4 & 0.6939 &    & 4 & 0.7564&     & 4 & 0.8026&      & 19& 0.7647&      & 14& 0.7866\\
   & 5 & 0.7009 &    & 5 & 0.7759&     & 5 & 0.8067&      & 20& 0.7551&      & 15& 0.7836\\
20*& 1 & 0.4477&     & 6 & 0.7640&     & 6 & 0.8056&   50*& 1 & 0.6509&      & 16& 0.7802\\
   & 2 & 0.6245&     & 7 & 0.7685&     & 7 & 0.7972&      & 2 & 0.7628&      &17 & 0.7772\\
   & 3 & 0.6845&     & 8 & 0.7630&     & 8 & 0.7970&      & 3 & 0.8207&      & 18& 0.7749\\
   & 4 & 0.7078&     & 9 & 0.7447&     & 9 & 0.7872&      & 4 & 0.8334&      & 19& 0.7755\\
   & 5 & 0.7277&     & 10& 0.7634&     & 10& 0.7838&      & 5 & 0.8392&      & 20& 0.7725\\
   & 6 & 0.7342&     &11 & 0.7598&     &11 & 0.7786&      & 6 & 0.8307&      & 21& 0.7696\\
   & 7 & 0.7382&     & 12& 0.7591&     & 12& 0.7732&      & 7 & 0.8308&      &22 & 0.7672\\
   & 8 & 0.7406&     & 13& 0.7553&     & 13& 0.7766&      & 8 & 0.8184&      & 23& 0.7722\\
   & 9 & 0.7321&     & 14& 0.7521&     & 14& 0.7732&      & 9 & 0.8142&      & 24& 0.7709\\
   & 10& 0.7259&     & 15& 0.7504&     & 15& 0.7662&      & 10& 0.8042&      & 25& 0.7660\\
   \hline
\end{tabular}
\begin{tabular}{ c c c c c c c c c c c c c c c }
&  &  &  &   &   & Normality &  &  &  &   &   & \\
\end{tabular}
\begin{tabular}{c c c c c c c c c c c c c c c }
\hline
 { $n$ }& $m$ & AP &  { $n$ }& $m$ & AP  &{ $n$ }& $m$ & AP &{ $n$ }& $m$ & AP&{ $n$ }& $m$ & AP\\
\hline
 10 & 1& 0.2765 & 30*  & 1& 0.5229 & 40* & 1 & 0.5776 & 40* & 16& 0.6098 & 50*  &11 & 0.6746\\
   & 2 & 0.3470 &     & 2 & 0.6154 &     & 2 & 0.6811 &     &17 & 0.5960 &      & 12& 0.6702\\
   & 3 & 0.3622 &     & 3 & 0.6547 &     & 3 & 0.7130 &     & 18& 0.5808 &      & 13& 0.6647\\
   & 4 & 0.3876 &     & 4 & 0.6628 &     & 4 & 0.7173 &     & 19& 0.5702 &      & 14& 0.6583\\
   & 5 & 0.3520 &     & 5 & 0.6559 &     & 5 & 0.7072 &     & 20& 0.5540 &      & 15& 0.6517\\
20*& 1 & 0.4276&      & 6& 0.6518 &     & 6 & 0.6995  &  50* & 1 & 0.6375 &      & 16& 0.6457\\
   & 2 & 0.5141&      & 7 & 0.6361 &     & 7 & 0.6911 &     & 2 & 0.7269  &      &17 & 0.6385\\
   & 3 & 0.5483&      & 8 & 0.6219 &     & 8 & 0.6786 &     & 3 & 0.7482  &      & 18& 0.6325\\
   & 4 & 0.5586&      & 9 & 0.6226 &     & 9 & 0.6689 &     & 4 & 0.7479  &      & 19& 0.6227\\
   & 5 & 0.5418&      & 10& 0.6104 &     & 10& 0.6598 &     & 5 & 0.7415  &      & 20& 0.6160\\
   & 6 & 0.5294&      &11 & 0.5983 &     &11 & 0.6521 &     & 6 & 0.7283  &     & 21& 0.6090\\
   & 7 & 0.5230&      & 12& 0.5799 &     & 12& 0.6425 &     & 7 & 0.7148  &     &22 & 0.5952\\
   & 8 & 0.5078&      & 13& 0.5631 &     & 13& 0.6363 &     & 8 & 0.7043  &     & 23& 0.5858\\
   & 9 & 0.4891&      & 14& 0.5505 &     & 14& 0.6258 &     & 9 & 0.6903  &     & 24& 0.5713\\
   & 10& 0.4744&      & 15& 0.5291 &     & 15& 0.6168 &     & 10& 0.6827  &     & 25& 0.5592\\
\hline
\end{tabular}\\
{*{\small $n=10r$ cases are observed by RSS scheme with 10 samples
and $r$ replication.}}
\end{table}

\begin{table}
  \centering
  \caption{Values of the window size $m$ with largest average of powers against
  alternatives}\label{optm}
\begin{tabular}{c c c} \hline
  & Optimal $m$(max average power) &   \\
 \cline{2-3}
 $n$  &  Exponentiality & Normality  \\
 \hline
 10  & 5(0.7009) &  4(0.3876)  \\
 20  & 8(0.7406) &  4(0.5586) \\
 30  & 5(0.7759) &  4(0.6628) \\
 40  & 5(0.8067) &  4(0.7173) \\
 50  & 5(0.8392) &  3(0.7482)  \\
   \hline
\end{tabular}

\end{table}

\newpage
\section*{References}
\begin{description}
\item Ahmad, I. A. and Lin, P. E.  ``A nonparametric
estimation of the entropy for absolutely continuous
distributions," \textit{IEEE Trans. Inf. Theory,} vol. 22,
372--375, 1976.

\item[{[}2{]}] Arizono, I. and  Ohta, H. ``A test for normality based on
Kullback-Leibler information," \textit{The American
Statistician,} vol. 43, pp. 20--23, 1989.

\item[{[}4{]}] Dmitriev,  Y. G. and  Tarasenko, F. P. ``On the
estimation of functional of the probability density and its
derivatives," \textit{Theory probab. Applic.,} vol. 18, 628--633,
1973.

\item[{[}5{]}] Ebrahimi, N.  and Habibullah, M.  ``Testing exponentiality
based on Kullback-Leibler information," \textit{J. Royal Statist.
Soc. B,} vol. 54, pp. 739--748, 1992.

\item[{[}7{]}]  Kullback, S. ``\textit{Information Theory and
Statistics,"} New York: Wiley, 1959.

\item[{[}8{]}]  Park, S. ``The entropy of consecutive order
statistics," \textit{IEEE Trans. Inform. Theory,} vol. 41, pp.
2003--2007, 1995.

\item[{[}9{]}]  Park, S. ``Testing exponentiality based on the
Kullback-Leibler information with the type II censored data,"
\textit{IEEE Trans. on Rel.,} vol. 54, pp. 22--26, 2005.

\item[{[}10{]}]  Shannon, C. E. ``A mathematical theory of
communications," \textit{Bell System Tech. J.,} vol. 27, pp.
379--423, 1948.

\item[{[}11{]}]  Teitler, S.,  Rajagopal, A. K. and  Ngai, K. L. ``Maximum
entropy and reliability distributions," \textit{IEEE Trans. on
Rel.,} vol. 35, pp. 391--395, 1986.

\item[{[}12{]}]  Vasicek, O. ``A test for normality based on sample
entropy," \textit{J. Royal Statist. Soc. B,} vol. 38, pp.
730--737, 1976.

\item  Arnold, B. C.; Balakrishnan, N. and Nagaraja, H. N.
(1992). {\em A First Course in Order Statistics}.\vspace{-1mm}
\item  Cochran, W. G. (1977). {\em Sampling Thechniques}. Third edition, Wiley, New York.\vspace{-1mm}
\item  Dell, T. R. and Clutter, J. L. (1972). Ranked set sampling theory with order statistics background. {\em Biometrics} {\bf 28}, 545 -- 555.\vspace{-1mm}
\item  McIntyre, G. A. (1952). A method of unbiased selective sampling, using ranked sets. {\em J. Agri. Res.} {\bf 3}, 385 -- 390.\vspace{-1mm}
\item  Muttlack, H. A. and L. I. McDonald (1990a). Ranked set sampling with respect to concomitant variables and with size biased probability of selection.
{\em Commun. Statist. -- Theory Meth.}, {\bf 19}, 205 --
219.\vspace{-1mm}
\item  Pearson, E. S. and Hartley, H. O. (1972). {\em Biometrika Tables for Statisticians}, Vol. 2. Cambridge University Press.\vspace{-1mm}
\item  Shorack, G. R. and Wellner, J. A. (1986). {\em Emperical Processes with Application to Statistics}, Wiley, New York.\vspace{-1mm}
\item  Stokes, S. L. (1977). Ranked set sampling with concomitant variables. {\em Commun. Statist. -- Theory Meth.}, {\bf 6}, 1207 -- 1212.\vspace{-1mm}
\item  Stokes, S. L. (1980a). Estimation of variance using judgment ordered ranked set samples. {\em Biometrics},
{\bf 36}, 35 -- 42.\vspace{-1mm}
\item  Stokes, S. L. (1980b). Inference on the correlation coefficient in bivariate normal populations from ranked set
samples. {\em J. Amer. Statist. Assoc.}, {\bf 75}, 989 --
995.\vspace{-1mm}
\item  Watterson, G.A. (1959). Linear estimation in censored samples from multivariate normal populations. {\em Ann. Math. Statist.} {\bf 30,} 814 -- 824.\vspace{-1mm}
\item  Yu, Philip, L. H. and Lam, K. (1997). Regression estimator in ranked set
sampling. {\em Biometrics}, {\bf 53}, 1070 -- 1080.\vspace{-1mm}

\end{description}
\end{document}